\documentclass[usenatbib]{mn2e}
\usepackage[totalwidth=480pt, totalheight=680pt]{geometry}
\usepackage{graphicx}
\usepackage{lineno}

\title{The Binary Nature of PSR J2032+4127}

\author[Lyne et al.]
	{A.~G.~Lyne$^{1,}$\thanks{E-mail: andrew.lyne@manchester.ac.uk}, 
        B.~W.~Stappers$^1$, M.~J.~Keith$^1$, P.~S.~Ray$^2$, M.~Kerr$^3$, F.~Camilo$^{4}$, 
\newauthor
        and T.~J.~Johnson$^{5}$
\\
	$^1$ Jodrell Bank Centre for Astrophysics, School of Physics and Astronomy,
	The University of Manchester, Manchester M13 9PL, UK
\\
        $^2$ Space Science Division, Naval Research Laboratory, 
        Washington, DC 20375-5352, USA
\\
        $^3$ CSIRO Astronomy and Space Science, Australia Telescope National
        Facility, PO Box 76, Epping, NSW 1710, Australia
\\
        $^4$ Columbia Astrophysics Laboratory, Columbia University, 
        New York, NY 10027, USA
\\
         $^5$ College of Science, George Mason University, Fairfax, VA 22030, 
         resident at Naval Research Laboratory, Washington, DC 20375, USA
\\}

\begin{document}

\date{}
\pagerange{\pageref{firstpage}--\pageref{lastpage}} \pubyear{2014}
\maketitle

\label{firstpage}
\begin{abstract}
PSR~J2032+4127 is a $\gamma$-ray and radio-emitting pulsar which has
been regarded as a young luminous isolated neutron star.  However, its
recent spin-down rate has extraordinarily increased by a factor of
two.  We present evidence that this is due to its motion as a member
of a highly-eccentric binary system with a $\sim15$-M$_\odot$ Be
star, MT91~213.  Timing observations show that, not only are the
positions of the two stars coincident within 0.4$''$, but timing
models of binary motion of the pulsar fit the data much better than a
model of a young isolated pulsar.  MT91~213, and hence the pulsar, lie
in the Cyg~OB2 stellar association, which is at a distance of only
1.4--1.7 kpc.  The pulsar is currently on the near side of, and
accelerating towards, the Be star, with an orbital period of 20--30 years.
The next periastron is well-constrained to occur in early 2018,
providing an opportunity to observe enhanced high-energy emission as
seen in other Be-star binary systems.
\end{abstract}


\section{Introduction}
PSR~J2032+4127 is a 143-ms pulsar discovered by the Large Area
Telescope (LAT) of the \textit{Fermi Gamma-ray Space Telescope}
\citep{aaa+09} and subsequently detected at radio frequencies
\citep{crr+09}.  Using $\gamma$-ray data as well as radio data, pulse
arrival-time analysis by those authors, and latterly by \cite{rkp+11},
showed that the pulsar had a large slow-down rate, indicating that it
was a young pulsar, a notion which has been supported by a recent
glitch in its rotation rate.  The timing analysis showed that the
projected position of the pulsar lay close to a V=11.95 Be
star, MT91~213 in the Cyg OB2 stellar association.  However,
\citet{crr+09} concluded that it was not a binary companion of
PSR~J2032+4127, based mainly upon the apparent lack of variations in
the pulsar rotation rate that would arise from the Doppler effects of
any reasonable (circular) orbital binary motion.  Subsequent studies
of the properties of the pulsar have all been based upon the
assumption that it is a solitary young energetic pulsar.  For
instance, \citet{aab+14} present a VERITAS detection of TeV~J2032+4130
and they discuss its likely association with PSR~J2032+4127, noting
that the extended nature of the source and the prevalence of pulsar
wind nebulae (PWNe) in the Galactic TeV source population argues that
it is a PWN powered by the pulsar. However they note that the extended
X-ray emission that is also spatially coincident with the source
\citep{hhs+07,bdb+06,mkk+11} is quite weak if it is from a PWN.  Six
years of both $\gamma$-ray and radio timing data are now available, and
we revisit the possibility that the pulsar is in orbit with MT91~213.

\section{Observations and Timing Noise Model}
\textit{Fermi} LAT data were used from shortly after the start of the
mission in August 2008 up to June 2014 (MJD 54682--56824).  Following
the maximum likelihood method of \citet{rkp+11}, we constructed
times-of-arrival (TOAs) with a 14-day cadence from ``reprocessed''
Pass 7 \textit{Fermi} LAT data.  For this analysis, we used
\texttt{SOURCE} class photons, excluding events with a zenith angle
$>100^\circ$ and when the spacecraft rocking angle exceeds $52^\circ$.
To improve sensitivity, we employed photon weighting using the
spectral models available in The Second \textit{Fermi} Large Area
Telescope Catalog of Gamma-ray Pulsars \citep{aaa+13}.

Radio timing observations were made with the NRAO Green Bank Telescope
(GBT) and the Lovell Telescope (LT) at Jodrell Bank.  The GBT
observations were primarily in bands centred on 820 MHz and 2000 MHz
during the early part of this period (MJD 54836--55589)
\citep{crr+09}, as well as MJD 56855--56857.  Radio observations
with the LT were made at approximately weekly intervals at around 1520
MHz for MJD 55222--56830.  As described by \citet{crr+09}, the
$\gamma$-ray and radio pulse profiles are very different in form.  In
order to establish the true alignment of the profiles, the dispersion
measure (DM) was initially determined using only radio data.
Inspection of the radio pulse profiles shows that there is little
change in shape between 800 MHz and 2000 MHz, and a single 
standard profile was used to obtain TOAs for all the radio
observations. The DM was determined at three epochs, using the three
800-MHz TOAs and their neighbouring 1520-MHz and 2000-MHz TOAs. This
allowed the time offset of the contemporaneous $\gamma$-ray TOAs to be
established using the {\sc jump} facility in {\sc tempo2}
\citep{hem06}. The mean value of these time offsets was consistent
with the alignment obtained by \citet{crr+09} and was applied to all
the $\gamma$-ray TOAs, allowing the delay between these TOAs and the
radio TOAs to be used to determine the DM throughout the data set.

The radio TOAs have errors of $\sim300\;\mu$s, about
half the errors of those provided by the \textit{Fermi} LAT.  In
total, about 400 good TOAs are available over six years (MJD
54682--56857).

\begin{figure}
\includegraphics[width=8.4cm, angle=0.0]{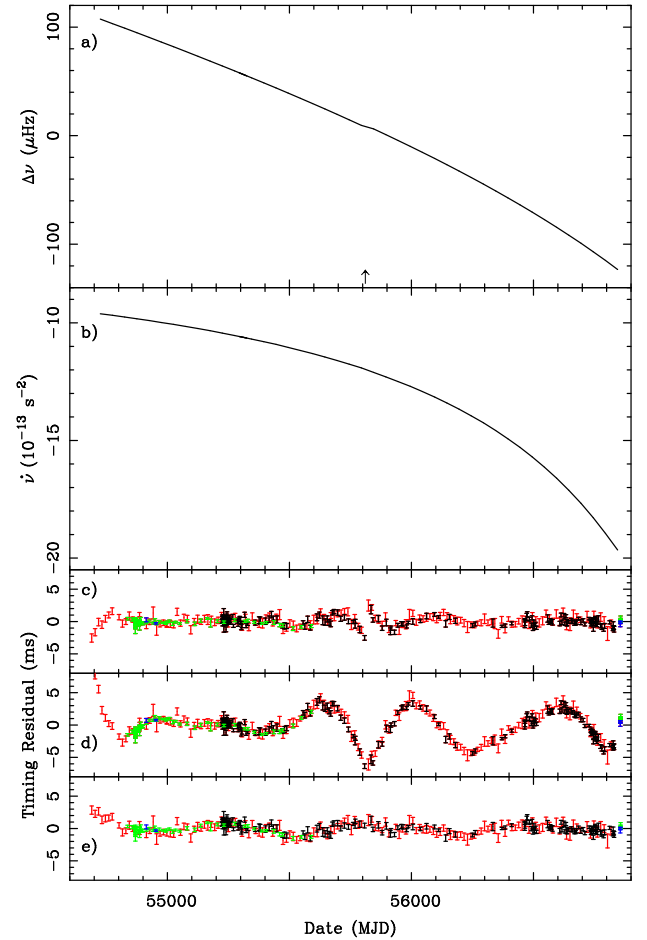}
\caption{The spin-frequency history of PSR~J2032+4127 determined over
six years.  a) The observed variation in the spin-frequency offset
$\Delta\nu$ from 6.98083 Hz, determined from fits to 150-day sets of
TOAs every 50 days, showing the monotonic spin-down of the
pulsar. This spin-down is interrupted briefly by a small glitch close
to MJD 55811, identified by an arrow. b) The variation of the
spin-frequency first derivative $\dot\nu$ determined over the same
time intervals as (a), showing a doubling of the magnitude.  The fits
for $\nu$ and $\dot\nu$ for (a) and (b) were conducted simultaneously
with the position fixed at $\alpha=20^{\rm h}32^{\rm m}13^{\rm s}.105$,
$\delta=41^\circ27'24''.36$. c) The timing residuals relative to the
7-derivative noise model presented in Table~1, column~2.  d) The
timing residuals relative to a 6-derivative noise model.  e) The
timing residuals relative to the best-fit ``binary'' model, which has
orbital period $P_{\rm b}$=8578 d and mass function $f_{\rm m}=10 \;
{\rm M}_\odot$ (Table~1, column~4).  In panels c, d and e, the colour
of the points indicates whether they were obtained from the LAT (red),
the LT (black), the GBT at 2000~MHz (green) or the GBT
at $\sim820$ MHz (blue).  All 3 timing models used in Figs 1c, 1d, 
and 1e do in fact include fits for the glitch parameters.  In the 
timing-noise models, there is significant covariance between glitch 
parameters and other polynomial parameters, so that some features of 
the glitch are absorbed in the polynomial parameters, leaving some 
sharp, unmodeled features of the glitch in the residuals.  It seems 
that the glitch and Keplerian parameters are not so covariant, and 
the glitch is fitted well. }
\label{residuals}
\end{figure}

\begin{figure}
 \includegraphics[width=9.0cm, angle=0.0]{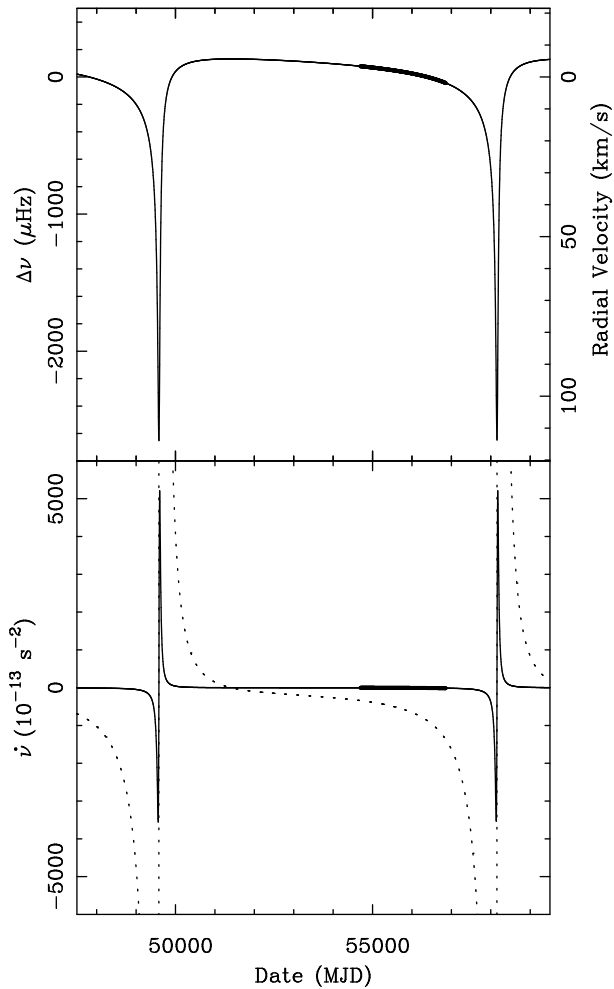} \caption{The
 orbital variation of the frequency $\nu$ and the first derivative
 $\dot\nu$ of PSR~J2032+4127 for the best-fit binary model, which has
 orbital period $P_{\rm b}$ = 8578 d and mass function $f_{\rm m} = 10
 \; {\rm M}_\odot$ (Table~1, column~4).  The scale on the right-hand
 axis of the upper panel indicates the magnitude of the orbital radial
 velocity.  The dotted line in the lower panel shows the variation in
 $\dot\nu$ on a scale expanded by a factor of 100.  The bold sections
 in the two plots indicate the range of the available data.}
\label{orbit}
\end{figure}

The pulsar shows substantial variations in rotation rate, indicating a
large amount of ``timing noise'', which is not unexpected in a young
pulsar (i.e. having a large intrinsic spin-down rate).  However, it
has recently become clear that the magnitude of these variations far
exceeds any previously encountered timing noise in a pulsar
(e.g. \citealt{hlk10}).  Fig.~1a presents the evolution of the
barycentric rotation frequency of the pulsar over the 6-year span of
the timing data, showing a substantial reduction in frequency during
this time.  Also apparent is an increase in the magnitude of the
slope, corresponding to an increase in the magnitude of the frequency
derivative, shown in Fig.~1b, by more than a factor of two, and a
decrease in the characteristic age by the same factor, confounding any
interpretation in terms of conventional pulsar spin-down. There is
also a barely-discernable minor glitch which occurred in Sept 2011
(MJD 55811) and which can be fitted by steps in frequency
($\Delta\nu$) and frequency derivative ($\Delta\dot{\nu}$). The whole
set of TOAs can be well-fitted by a model involving the position,
three glitch parameters, the rotation frequency and 7 derivatives
(Table~1, column 2).  The timing residuals (the difference between the
observed and model-based TOAs) relative to this model, which we will
henceforth refer to as the ``noise model'', are shown in Fig.~1c.

The behaviour in the rotation seen in Figs.~1a and 1b has never been
seen in any other isolated pulsar, and is not characteristic of either
a post-glitch recovery, during which the magnitude of the slow-down
rate is usually observed to decrease (e.g. \citealt{elsk11}), or of
timing noise, which is usually characterised by switches, often
quasi-periodic, between values of slow-down rate rather than the
smooth variation seen here \citep{lhk+10,lyn13}.  Most anomalous
of all is the doubling of the slow-down rate, also not seen in any
other isolated pulsar.

\begin{table*}
\caption{Model fits to the TOAs of PSR~J2032+4127.  Observed values of
parameters for a 7-derivative polynomial ``Noise Model'' fit, and for
two best-fit ``Binary Models'', constrained to have mass functions
$f_{\rm m}=10$ and 20 M$_\odot$ respectively and with
errors determined from a Monte-Carlo analysis. One-$\sigma$ uncertainties are
given in parentheses after the values and are in units of the least
significant quoted digit.}
\label{spinpars}
\begin{tabular}{lllll}
\hline \hline
\multicolumn{1}{c}{Parameter} & \multicolumn{1}{c}{Noise Model}    &
\multicolumn{1}{c}{Parameter} & \multicolumn{1}{c}{Binary $f_{\rm m}=10{\rm M}_\odot$} &\multicolumn{1}{c}{Binary $f_{\rm m}=20{\rm M}_\odot$} \\
\hline
Right Ascension, $\alpha$ (J2000.0)  & $20^{\rm h}32^{\rm m}13.^{\rm s}111(8)$ &  & $20^{\rm h}32^{\rm m}13.^{\rm s}107(8)$  & $20^{\rm h}32^{\rm m}13.^{\rm s}105(8)$\\
Declination, $\delta$ (J2000.0)      & $41^\circ 27'24''.17(10)$ &  & $41^\circ 27'24''.36(10)$  & $41^\circ 27'24''.36(10)$ \\
\\
Epoch of frequency, $t_{\rm 0}$ (MJD)                     & 55714.0    &  & 55700.0    & 55700.0 \\
Freq, $\nu_0$ (Hz)                                  & 6.9808479486(10) &  & 6.980808(12)  & 6.980806(12) \\
Freq 1st deriv, $\dot{\nu}_0$ ($10^{-12}$s$^{-2}$)  & $-$1.16583(12)  &  & $-$0.61(5)  & $-$0.61(5) \\
\\
Glitch Epoch, $T_{\rm g}$ (MJD)                     & 55810.51(2)&  & 55810.76(7) & 55810.76(7)\\
Freq, $\Delta\nu_{\rm g}$ ($10^{-6}$ Hz)            & 1.907(2)  &  & 1.907(2)   & 1.907(2)  \\
Freq 1st deriv, $\Delta\dot{\nu}_{\rm g}$ ($10^{-16}$s$^{-2}$)&  $-$4(3) &  &  $-$10(2)   &  $-$10(2)  \\
\\
DM (pc cm$^{-3}$)                                   & 114.66(3) &  & 114.65(3)  & 114.65(3) \\
DM deriv, DM1 (pc cm$^{-3}$y$^{-1}$)                & 0.00(1) &  & 0.00(1)  & 0.00(1) \\
\\
Timing noise parameters:                            &              & Binary parameters: \\
Freq 2nd deriv, $\ddot{\nu}_0$ ($10^{-20}$s$^{-3}$) & $-$0.3651(7) & Orbital period, $P_{\rm b}$ (d)   & 8578(1200)  & 8625(1200) \\
Freq 3rd deriv, $\nu_0^{(3)}$ ($10^{-28}$s$^{-4}$)  & $-$0.439(2)  & Epoch of Periastron, $T_{\rm 0}$ (MJD)  & 58161(31)   & 58153(31) \\
Freq 4th deriv, $\nu_0^{(4)}$ ($10^{-36}$s$^{-5}$)  & $-$0.690(14) & Proj. semi-major axis, $x$ (lt-s) & 8819(1250)  & 11151(1250) \\
Freq 5th deriv, $\nu_0^{(5)}$ ($10^{-44}$s$^{-6}$)  & $-$1.26(9)   & Eccentricity, $e$                 & 0.93(3)     & 0.95(3) \\
Freq 6th deriv, $\nu_0^{(6)}$ ($10^{-52}$s$^{-7}$)  & $-$6.2(3)    & Long. of Periastron $\omega$ (deg)& 12(4)       & 10(4) \\
Freq 7th deriv, $\nu_0^{(7)}$ ($10^{-60}$s$^{-8}$)  & $-$27(3)     &  &  \\
\\
Rms timing residual, $\sigma$ (ms)                  & 0.53         &  & 0.57  & 0.56 \\
\\
\hline
\end{tabular}
\end{table*}

\section{A binary Model}
We believe that the only plausible origin of such a large variation in
the observed slow-down rate of a pulsar lies in the Doppler effects of
binary motion with another star.  While we note the remark by
\cite{crr+09} that there is no evidence of any short-period binary
motion and that the binary period (of any circular orbit) must be in
excess of 100 years, we have explored the possibility that 
the pulsar is actually a member of a long-period binary system with a
large orbital eccentricity.

First, we sought fits of binary models to the rotation-frequency data
of Fig.~1a.  While good fits to the data were possible using models
for eccentric binary orbits with orbital periods $P_{\rm b}$ in excess
of about 6000 days (16 years), it soon became clear that there were
strong covariances between some of the fitted parameters, arising from
the small orbital phase range of the available data.  In particular,
there were large covariances between the intrinsic pulsar slow-down
rate, $\dot\nu$, the orbital period $P_{\rm b}$ and the projected
semi-major axis of the orbit $x=(a/c)\;{\rm sin}\; i$, where $a$
is the semi-major axis of the orbit, $i$ is the inclination of the
plane of the orbit to the plane of the sky and $c$ is the speed of
light. These covariances allowed many good, but not unique, fits to
the data.  We therefore explored fits to the data for a range of fixed
values of $P_{\rm b}$ between 6000 and 12000 days at 500-day
intervals, and, for each of these, a range of fixed values of $x$,
corresponding to fixed values of mass function $f_{\rm m}$ of 2, 5, 10
and 20 ${\rm M}_\odot$. $f_{\rm m}$ is a function of the masses of the
neutron star ($M_{\rm p}$) and its companion ($M_{\rm c}$) and orbital
inclination $i$ and is determined from $P_{\rm b}$ and $x$ from
Kepler's laws by:
\begin{equation}
f_{\rm m} = \frac{(M_{\rm c}\; {\rm sin}\; i)^3}{(M_{\rm p}+M_{\rm c})^2} =
\frac{4\pi^2}{G} \frac{x^3}{P_{\rm b}^2\;\;}
\label{f_m}
\end{equation}
where $G$ is Newton's gravitational constant.  If $M_{\rm p}, \;M_{\rm
c}$ and $f_{\rm m}$ are in solar masses, $x$ is in light-sec and $P_{\rm
b}$ in days, then
\begin{equation}
x=9.766 \sqrt[3]{f_{\rm m} P_{\rm b}^2}.
\label{x}
\end{equation}
Fig.~2 shows an example of one of the best fits to the data, for
$P_{\rm b}$ = 8578 days, $f_{\rm m}$ = 10 M$_\odot$.  Note that the
data span occupies only about 20\% of the orbit.  Remarkably, the root
mean square (rms) of the frequency residuals (the differences
between the measured and model frequency values) for this simple
model was only approximately $10^{-5}$ of the total frequency
variation during this time and is consistent with the measurement
errors.

\begin{figure}
 \includegraphics[width=11.0cm, angle=0.0]{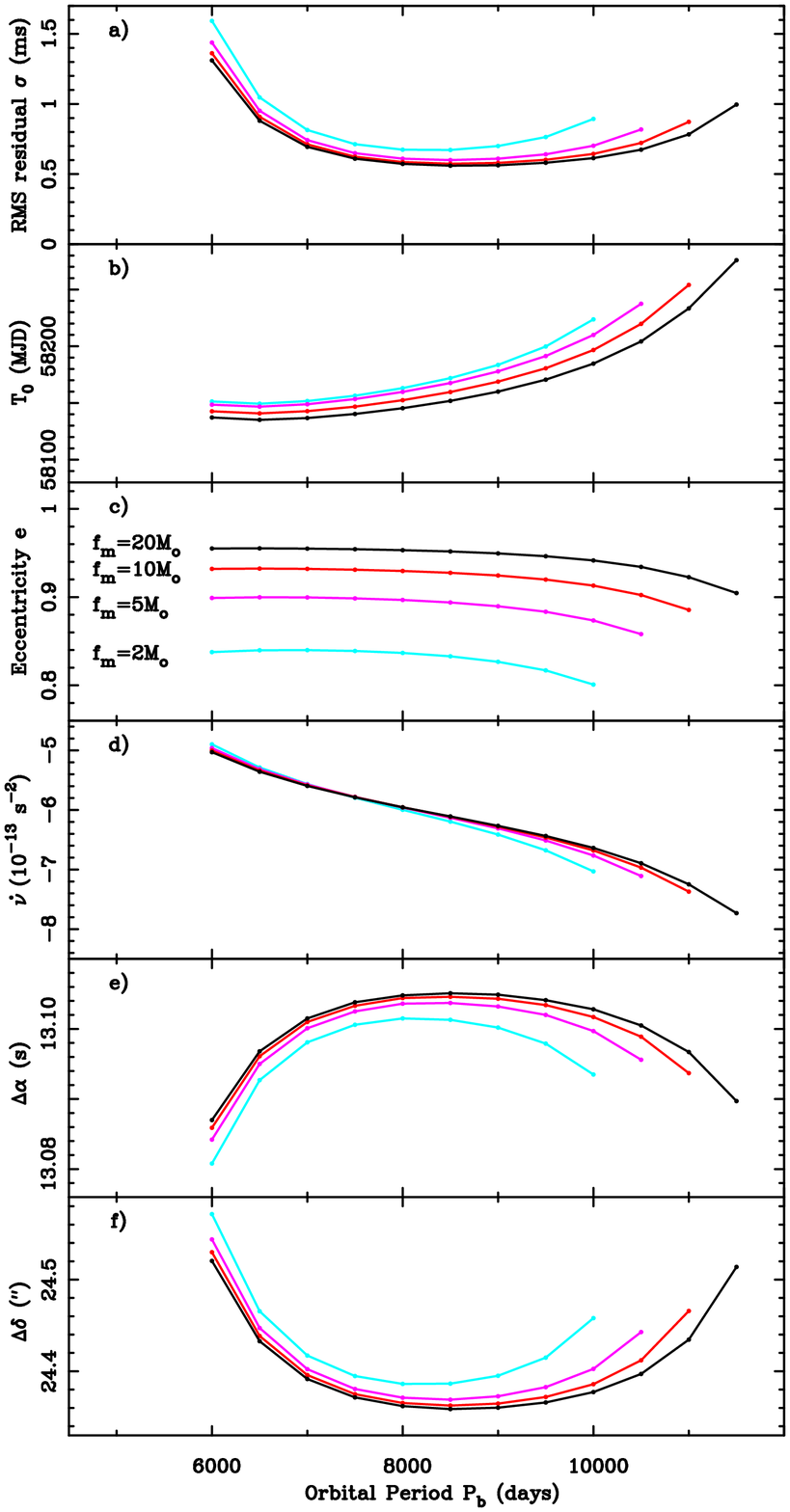} \caption{The
 results of fits to the TOAs of a number of binary models having
 different fixed values of orbital period $P_{\rm b}$ and mass
 function $f_{\rm m}$.  a) The rms timing residual ($\sigma$) with the
 colour-coding the same as that defined in (c). b) The epoch of
 periastron ($T_{\rm 0}$). c) The orbital eccentricity ($e$). d) The
 pulsar rotational frequency derivative ($\dot\nu$). e) The right
 ascension of the pulsar less $20^{\rm h}32^{\rm m}$ ($\Delta\alpha$).
 f) The declination less $41^\circ27'$ ($\Delta\delta$).}
\label{fits}
\end{figure}

\begin{figure}
\includegraphics[width=8.0cm, angle=0.0]{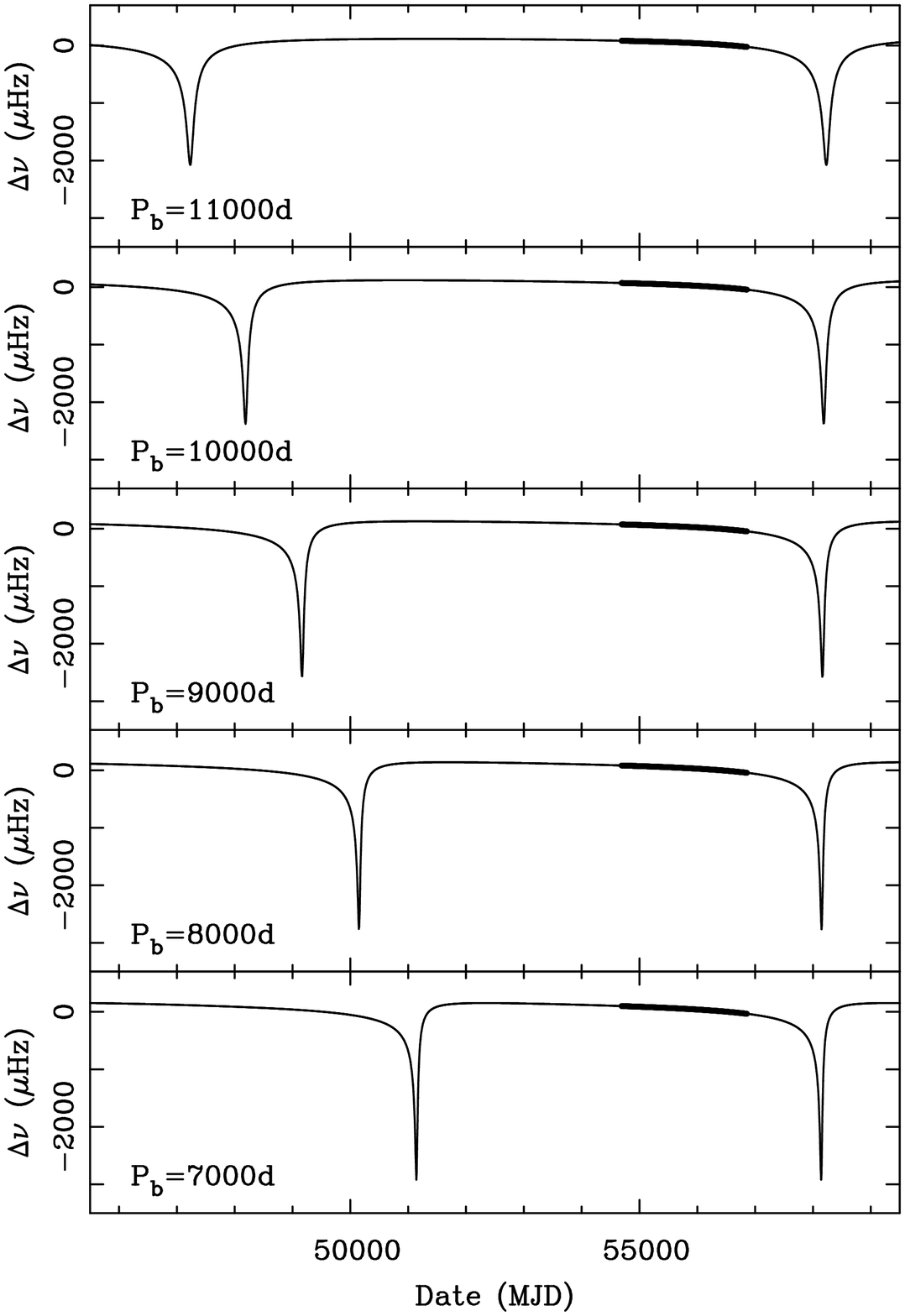} 
\caption{The orbital variation of the frequency $\nu$ of
PSR~J2032+4127 for the best fit binary models, for the indicated
orbital periods P$_{\rm b}$ and having mass function $f_{\rm m}=10$
M$_\odot$.  Note that the epochs of the next periastron around 58150
are almost independent of the value of $P_{\rm b}$.  The bold sections
in the plots indicate the range of the available data.}
\label{orbits}
\end{figure}

For each $P_{\rm b}$, $f_{\rm m}$ pair, the ephemeris resulting from
the frequency fit was subsequently used as the basis of a coherent
timing analysis of the TOAs using {\sc tempo2}.  Keeping $P_{\rm b}$
and $x$ at their fixed values, the three other Keplerian parameters,
the pulsar position, DM and its first derivative, DM1, the rotation
frequency and its first derivative and three glitch parameters were
all fitted to the TOAs.  Fig.~3 summarises the results of the fits.
In particular, Fig.~3a shows the rms of the timing residuals relative
to the fitted model as a function of $P_{\rm b}$ and $f_{\rm m}$.
There is a broad minimum in the rms with a value of about 0.6 ms for
7500 d $<$ $P_{\rm b}$ $<$ 9500 d and all values of $f_{\rm m}
\geq 5 \; {\rm M}_\odot$ provide indistinguishably good fits.  
Best-fit models were obtained for $f_{\rm m}=10 \; {\rm M}_\odot$ and
$f_{\rm m}=20 \; {\rm M}_\odot$ by including a fit for $P_{\rm b}$ and
the results are given in Table~1, columns 4 and 5.  Because of the
strong covariance between some of the parameters, we used a
Monte-Carlo method to estimate the errors in the fitted parameters
that are given in the table.  The Monte-Carlo analysis used
simulations of the pulsar observations seeded with the best-fit
parameter file.  The simulated data sets had the same epochs of
the combined LAT and radio TOAs, giving the same cadence and
orbital coverage as the real data. A noise source was added to the
data, defined by the power spectral density:
\begin{equation}
P(f) = P_w + A \left(1 + \left(\frac{f}{f_c}\right)^2\right)^{\alpha/2},
\end{equation}
where the white noise, $P_w = 6.5\times10^{-24}\,$yr$^{3}$, and the
power-law parameters are $A=9\times10^{-20}\,$yr$^{3}$, $f_c =
0.2\,$yr$^{-1}$ and $\alpha = -4$. The noise parameters were chosen to
match the observed noise spectrum. The {\sc toasim} plugins from {\sc
tempo2} were used to generate $10^4$ realisations of the noise and the
full fitting process was applied to each realisation. The error
estimates were determined from the variance of the resultant fit
parameters.

Fig.~1e shows the timing residuals for the binary fit for $f_{\rm
m}=10 \; {\rm M}_\odot$ given in Table~1, column~4, showing that the
rms is almost entirely limited by measurement errors, with any
remaining unmodelled systematic trends at a level that is comparable
with those from pulsars of a similar age.  We note that these binary
models involve the fitting of 14 free parameters to achieve an rms
residual of 0.57 ms.  This is one parameter fewer than required for
the timing-noise model with similar rms residuals that is presented
in Fig.~1c and Table~1, and based upon a 7-derivative polynomial
model, achieving an rms residual of 0.53 ms.  Fig.~1d shows the
residuals based upon a 6-derivative timing-noise model that has the
same number of fitted parameters as the binary model and is clearly a
poorer description of the data.

We conclude that there is a range of simple binary models that
describe the TOA data very well.  It is also clear that the reason we
cannot define the orbit more precisely at present is that, because the
data span covers only about 1/5 of an orbit, the differences between
the models are buried within the measurement errors of the TOAs and
likely timing noise.

However, Figs.~3b--3f show that the values of many of the fitted
parameters do not change much near the rms residual minimum.  For all
the models near this minimum (Fig.~3a), the eccentricity increases
from about 0.90 to 0.95 with increasing $f_{\rm m}$
(Fig. 3c). However, the corresponding $T_{\rm 0}$ lies in a restricted
range between MJD 58150 and 58200 (Figs.~3b and 4).  Thus the next
periastron passage will occur sometime during 2018 February or March.
The intrinsic frequency derivative for all these models is about
$-6(1)\times10^{-13}$ s$^{-2}$ (Fig. 3d), corresponding to a
characteristic age $\tau=\nu/2\dot\nu$ of about 180 kyr, which is
substantially greater than the (rapidly decreasing!) characteristic
age of the pulsar in the young-isolated-pulsar model, which has
changed from $\sim120$~kyr to $\sim55$~kyr over the six years.  The
values of right ascension $\alpha$ and declination $\delta$ are not
strongly model-dependent (Figs. 3e and 3f) and are about $20^{\rm
h}32^{\rm m}13^{\rm s}.106(8)$ and $+41^\circ27'24''.36(10)$, consistent
with the position of the Be star, MT91 213, which is at $20^{\rm
h}32^{\rm m}13^{\rm s}.137(18)$ and
$+41^\circ27'24''.48(20)$\footnote{Owing to a typographical error,
\citet{crr+09} listed the last digits of declination of MT91~213 as
24.28; the correct value is 24.48 as used here.}, and which
\citet{crr+09} considered, but rejected, as a possible companion of
the pulsar.  The conflicting conclusion of that study with this paper
arose from their assumption of a circular orbit and the happenstance
that their observations were conducted near apastron, where the
gravitational influence of the companion on the pulsar is small, so
that Doppler effects result in a small value of the pulsar second
rotational frequency derivative.

\section{Discussion}
If MT91~213 is indeed the companion star, what additional constraints
might be placed upon the system?  \cite{kkk+07} estimate the mass of
the star to be $M_{\rm c}\sim17.5$ M$_\odot$, based upon their
estimate of its spectral type (B0V) and the study by \citet{msh05} who
suggest that such spectrally-determined mass estimates may be in error
by 35--50\%. However, more recently \citet{hns10} have estimated the
mass of B0V stars to be $15.0\pm2.8$ M$_\odot$ using 2MASS
spectrometry and Hipparchos parallax data.  We adopt this latter value
in the following discussion.  Unfortunately, the inclination of the
plane of the orbit to the plane of the sky is unknown.  For random
orientations of a binary orbit, the median of the probability density
function of inclination $i$ occurs at $i=60^\circ$.  For an orbit with
this inclination and a pulsar of mass $M_{\rm p} = 1.35 \; {\rm
M}_\odot$, the mass function (Equation \ref{f_m}) would thus be
$f_{\rm m} = 8.2 \; {\rm M}_\odot$, which corresponds approximately to
the first binary model in Table~1, column~4, and the orange lines in
Fig.~3.  The greatest possible value of the mass function for these two
stars is about 15 M$_\odot$ which occurs if $i=90^\circ$.  There is no
doubt that an orbit with a mass function of $\sim5-15$ M$_\odot$,
having an orbital period lying between 7000 and 11000 days, or about
20--30 years, is entirely consistent with the data (Fig.~4).  Although
larger mass functions also provide satisfactory fits to the data, they
are inconsistent with a companion that has the mass of MT91~213.

\begin{figure*}
\includegraphics[width=11.5cm, angle=-90.0]{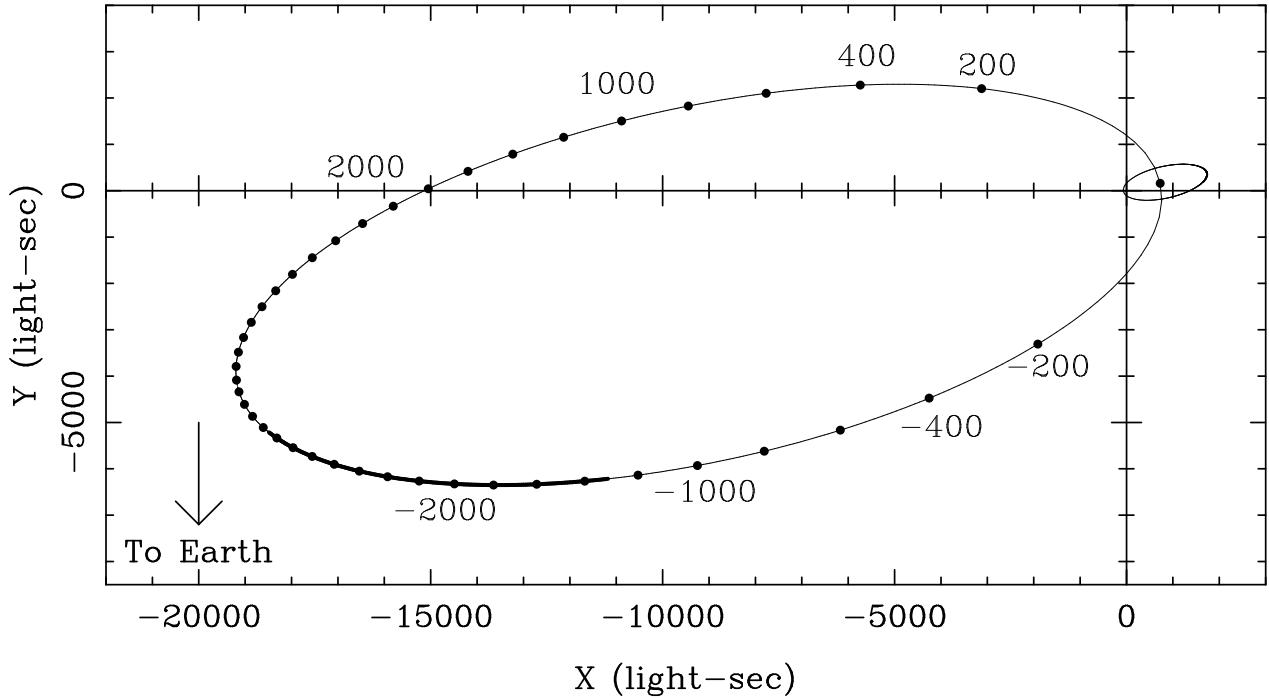} 
\caption{Schematic diagram showing the approximate orbital motions of
PSR~J2032+4127 and its Be-star companion MT91~213 about their common
centre of mass for the binary model given in Table~1, column~4 for
$f_{\rm m}=10 \; {\rm M}_\odot$, projected onto the plane containing
the line-of-sight and the major axis of the orbit.  The inclination
$i$ of the plane of the orbit to the plane of the sky is assumed to be
60$^\circ$.  The markers are at 200-day intervals and indicate the time
from periastron.  The thick line shows the portion of the pulsar orbit
covered by the observations reported here. The pulsar is moving
counter-clockwise in the diagram and is currently
(MJD~56912) on the near side of the Be star and about 1300 days
before periastron passage.  Note that the orbital velocity is
proportional to the separation of the markers, a 1000-light-sec
separation indicating a velocity of about 18 km s$^{-1}$.  The small
ellipse near the origin shows the orbit of the Be star, assuming that
it has a mass of 15~M$_\odot$ and that the pulsar has a mass of
1.35~M$_\odot$.  At an estimated distance of 1.7~kpc, the projected
major axis of the pulsar orbit will subtend an angle of about 25~mas.}
\label{orbit_plan}
\end{figure*}

Two other pulsar/Be-star binary systems are known,
PSR~B1259$-$63/SS~2883 \citep{jml+92} and PSR~J1638$-$4725
\citep{lfl+06,lyn08}, having orbital periods $P_{\rm b}$ of 1237 d and
1941 d respectively, and companion masses $M_{\rm c}$ of $\sim 4 \;
{\rm M}_\odot$ and $\sim 8 \; {\rm M}_\odot$.  The
PSR~J2032+4127/MT91~213 system is more extreme than either of these,
with $P_{\rm b}\sim9000$ d and $M_{\rm c}\sim 15$ M$_\odot$.  Fig.~5
shows the most likely orbital configuration. The pulsar is presently
on the Earth side of the Be star and is beginning to move rapidly in
towards periastron, after which it will move behind the star and any
circumstellar disk.  Such stars have substantial stellar winds and, in
the cases of both PSR~B1259$-$63 and PSR~J1638$-$4725, as well as
radio eclipses, there are increases in both dispersion measure (DM)
and multi-path scattering close to periastron, which arise as the
pulsar becomes more embedded in the dense circumstellar environment.
No significant change has yet been detected in the scattering or DM of
PSR~J2032+4127, with the fitted values of its first derivative DM1
consistent with zero (Table~1), but it is still far from periastron
and this is not unexpected.  Following periastron in early 2018, the
pulsar will move behind MT91~213 and may be eclipsed, at least in the
radio, by its atmosphere or disk, depending upon the inclination of
the orbit to the line-of-sight.

Such a large, long-period system is unique, but its existence and its
survival of the velocity kick that the pulsar probably experienced when 
it was formed is perhaps not surprising, because of the deep
gravitational well of the $\sim15$-M$_\odot$ companion.  The separation
of the markers in Fig.~5 gives an indication of the orbital velocity,
showing that, although the velocity at apastron is only a few km
s$^{-1}$, at periastron the velocity is in excess of 100 km s$^{-1}$.
It is a system that, unlike the majority of pulsars, did not quite
become unbound after the neutron star formation.

In all the allowed binary models, periastron will occur in just over
three years time.  Much of the degeneracy present in the fits will be
resolved as periastron is approached.  In particular, the rapid
increase in slow-down rate during the next 1--2 years will establish
the magnitude of the eccentricity and hence the stellar separation at
periastron.  All evidence indicates that MT91~213 is in the Cyg~OB2
stellar association, which is located at a distance of 1.4--1.7~kpc,
determined both by spectroscopic parallax and by VLBI determination of
the trigonometric parallax of masers associated with the association
stars \citep{mt91,han03,rbs+12}; in the binary hypothesis, this is
therefore also the distance of PSR~J2032+4127.  We note that, at this
distance, the projected major axis of the pulsar binary orbit will
subtend about 25~mas viewed from the Earth.  It should be possible to
track the pulsar around the orbit using a VLBI network such as the
VLBA or the EVN, hence establishing the size and the projected
eccentricity of the orbit.  Together with timing measurements, such
observations permit the direct determination of the inclination $i$ of
the orbit to the plane of the sky and the distance of the system from
the Earth.

The spin-down luminosity of PSR~J2032+4127 in the binary models
presented here is $\dot{E}= 1.7\times 10^{35}$\,erg\,s$^{-1}$, 62\% of
the \cite{crr+09} isolated-pulsar value. The efficiency in converting
rotational kinetic energy to $\gamma$-ray luminosity $L_{\gamma}$,
$\eta \equiv L_{\gamma}/\dot E$, is therefore increased by a factor of
1.6.  Using the distance of 1.4--1.7~kpc and the updated $L_{\gamma}$
from \cite{aaa+13}, and assuming as usual an isotropic $\gamma$-ray
beam, then $\eta =$15\%--20\%, which is unremarkable for a young
pulsar (the much higher value of $\eta$ presented in \citet{aaa+13}
uses the DM-based distance of 3.7\,kpc).

We also note that this system is still fairly young, the pulsar having
a characteristic age of 180~kyr and the Be star having a total
lifetime of only a few million years, which needs to encompass the
lifetimes of the pulsar progenitor star and of the pulsar itself.

The fact that PSR~J2032+4127 is a member of a long period binary
system with a massive companion star means that we should revisit the
interpretation of the spatially coincident extended TeV $\gamma$-ray and
X-ray emission \citep{aab+14}. \citet{kpd+14} have recently shown that
the similar system PSR~B1259$-$63 exhibits variable extended X-ray
emission that is located outside the binary system and is moving
away. It may be that this is emission that is associated with a
circum-binary shock which can be generated between the winds of a
pulsar and a massive star \citep{bbk+12}.  We note also that if the
interpretation of the TeV emission extension is to do with the proper
motion of the binary then the increased age will result in a reduced
transverse velocity of less than 30\,km\,s$^{-1}$ compared to that
derived by \citet{aab+14}.

In conclusion, there is a restricted range of binary orbits that
describe the observed timing data very well and explain the
extraordinary and unique growth in the observed rate of rotational
slow-down.  This and the coincidence of the precise positions of the
pulsar and MT91~213 all point to the two stars comprising a
neutron-star/Be-star binary system in the Cyg~OB2 association.  Near
periastron, the pulsar may well become obscured in the radio by
free-free absorption and by severe pulse scattering in the Be-star
circumstellar wind and disk as it passes behind the star, in the same
way as the Be star SS~2883 obscures PSR~B1259$-$63 at periastron
\citep{jml+96}.  However, studies of PSR~J2032+4127's DM and rotation
measure (RM) variations are likely to present a rare opportunity to
study the density of the stellar wind and any circumstellar disk and
the magnetic field of the Be star.  Although it may become obscured in
the radio for a short while, it should be possible to track its
rotation and orbit around periastron by observation in
$\gamma$ rays. Close to periastron, the MT91~213 Be star may also
display optical and/or X-ray variability as a direct result of its
interaction with the pulsar. Even now MT91~213 displays X-ray
variability (e.g.~\citealt{rnw+14}) and optical variability
(e.g.~\citealt{sab14}; see also \citealt{crr+09}), although these are
not atypical of the intrinsic variability observed in other Be stars
and almost surely have nothing to do with the neutron star.

\section*{Acknowledgments}
The \textit{Fermi} LAT Collaboration acknowledges generous ongoing
support from a number of agencies and institutes that have supported
both the development and the operation of the LAT as well as
scientific data analysis.  These include the National Aeronautics and
Space Administration and the Department of Energy in the United
States, the Commissariat \`a l'Energie Atomique and the Centre
National de la Recherche Scientifique / Institut National de Physique
Nucl\'eaire et de Physique des Particules in France, the Agenzia
Spaziale Italiana and the Istituto Nazionale di Fisica Nucleare in
Italy, the Ministry of Education, Culture, Sports, Science and
Technology (MEXT), High Energy Accelerator Research Organization (KEK)
and Japan Aerospace Exploration Agency (JAXA) in Japan, and the
K.~A.~Wallenberg Foundation, the Swedish Research Council and the
Swedish National Space Board in Sweden.  Additional support for
science analysis during the operations phase is gratefully
acknowledged from the Istituto Nazionale di Astrofisica in Italy and
the Centre National d'\'Etudes Spatiales in France.

Pulsar research at JBCA is supported by a Consolidated Grant from the UK 
Science and Technology Facilities Council (STFC). 

The GBT is operated by the National Radio Astronomy Observatory, a
facility of the National Science Foundation operated under cooperative
agreement by Associated Universities, Inc.

We are grateful to Jules Halpern for illuminating discussions.


\begin{thebibliography}{27}
\expandafter\ifx\csname natexlab\endcsname\relax\def\natexlab#1{#1}\fi

\bibitem[{{Abdo} {et~al}\mbox{.}(2009){Abdo}, {Ackermann}, {Ajello},
  {Anderson}, {Atwood}, {Axelsson}, {Baldini}, {Ballet}, {Barbiellini},
  {Baring}, {Bastieri}, {Baughman}, {Bechtol}, {Bellazzini}, {Berenji},
  {Bignami}, {Blandford}, {Bloom}, {Bonamente}, {Borgland}, {Bregeon}, {Brez},
  {Brigida}, {Bruel}, {Burnett}, {Caliandro}, {Cameron}, {Caraveo},
  {Casandjian}, {Cecchi}, {{\c C}elik}, {Chekhtman}, {Cheung}, {Chiang},
  {Ciprini}, {Claus}, {Cohen-Tanugi}, {Conrad}, {Cutini}, {Dermer}, {de
  Angelis}, {de Luca}, {de Palma}, {Digel}, {Dormody}, {do Couto e Silva},
  {Drell}, {Dubois}, {Dumora}, {Farnier}, {Favuzzi}, {Fegan}, {Fukazawa},
  {Funk}, {Fusco}, {Gargano}, {Gasparrini}, {Gehrels}, {Germani}, {Giebels},
  {Giglietto}, {Giommi}, {Giordano}, {Glanzman}, {Godfrey}, {Grenier},
  {Grondin}, {Grove}, {Guillemot}, {Guiriec}, {Gwon}, {Hanabata}, {Harding},
  {Hayashida}, {Hays}, {Hughes}, {J{\'o}hannesson}, {Johnson}, {Johnson},
  {Johnson}, {Kamae}, {Katagiri}, {Kataoka}, {Kawai}, {Kerr}, {Kn{\"o}dlseder},
  {Kocian}, {Kuss}, {Lande}, {Latronico}, {Lemoine-Goumard}, {Longo},
  {Loparco}, {Lott}, {Lovellette}, {Lubrano}, {Madejski}, {Makeev}, {Marelli},
  {Mazziotta}, {McConville}, {McEnery}, {Meurer}, {Michelson}, {Mitthumsiri},
  {Mizuno}, {Monte}, {Monzani}, {Morselli}, {Moskalenko}, {Murgia}, {Nolan},
  {Norris}, {Nuss}, {Ohsugi}, {Omodei}, {Orlando}, {Ormes}, {Paneque},
  {Parent}, {Pelassa}, {Pepe}, {Pesce-Rollins}, {Pierbattista}, {Piron},
  {Porter}, {Primack}, {Rain{\`o}}, {Rando}, {Ray}, {Razzano}, {Rea}, {Reimer},
  {Reimer}, {Reposeur}, {Ritz}, {Rochester}, {Rodriguez}, {Romani}, {Ryde},
  {Sadrozinski}, {Sanchez}, {Sander}, {Parkinson}, {Scargle}, {Sgr{\`o}},
  {Siskind}, {Smith}, {Smith}, {Spandre}, {Spinelli}, {Starck}, {Strickman},
  {Suson}, {Tajima}, {Takahashi}, {Takahashi}, {Tanaka}, {Thayer}, {Thompson},
  {Tibaldo}, {Tibolla}, {Torres}, {Tosti}, {Tramacere}, {Uchiyama}, {Usher},
  {Van Etten}, {Vasileiou}, {Vilchez}, {Vitale}, {Waite}, {Wang}, {Watters},
  {Winer}, {Wolff}, {Wood}, {Ylinen}, {Ziegler}, \& {Fermi LAT
  Collaboration}}]{aaa+09}
{Abdo} A.~A. {et~al.}, 2009, Science, 325, 840

\bibitem[{{Abdo} {et~al}\mbox{.}(2013){Abdo}, {Ajello}, {Allafort}, {Baldini},
  {Ballet}, {Barbiellini}, {Baring}, {Bastieri}, {Belfiore}, {Bellazzini}, \&
  et~al.}]{aaa+13}
{Abdo} A.~A. {et~al.}, 2013, ApJS, 208, 17

\bibitem[{{Aliu} {et~al}\mbox{.}(2014){Aliu}, {Aune}, {Behera}, {Beilicke},
  {Benbow}, {Berger}, {Bird}, {Buckley}, {Bugaev}, {Cardenzana}, {Cerruti},
  {Chen}, {Ciupik}, {Connolly}, {Cui}, {Duke}, {Dumm}, {Errando}, {Falcone},
  {Federici}, {Feng}, {Finley}, {Fortin}, {Fortson}, {Furniss}, {Galante},
  {Gillanders}, {Griffin}, {Griffiths}, {Grube}, {Gyuk}, {Hanna}, {Holder},
  {Hughes}, {Humensky}, {Kaaret}, {Kargaltsev}, {Kertzman}, {Khassen}, {Kieda},
  {Krawczynski}, {Lang}, {Madhavan}, {Maier}, {Majumdar}, {McCann}, {Moriarty},
  {Mukherjee}, {Nieto}, {O'Faol{\'a}in de Bhr{\'o}ithe}, {Ong}, {Otte},
  {Pandel}, {Perkins}, {Pohl}, {Popkow}, {Prokoph}, {Quinn}, {Ragan},
  {Rajotte}, {Reyes}, {Reynolds}, {Richards}, {Roache}, {Sembroski}, {Skole},
  {Staszak}, {Telezhinsky}, {Theiling}, {Tucci}, {Tyler}, {Varlotta},
  {Vincent}, {Wakely}, {Weekes}, {Weinstein}, {Welsing}, {Williams}, \&
  {Zitzer}}]{aab+14}
{Aliu} E. {et~al.}, 2014, ApJ, 783, 16

\bibitem[{{Bosch-Ramon} {et~al}\mbox{.}(2012){Bosch-Ramon}, {Barkov},
  {Khangulyan}, \& {Perucho}}]{bbk+12}
{Bosch-Ramon} V., {Barkov} M.~V., {Khangulyan} D., {Perucho} M., 2012, A\&A,
  544, A59

\bibitem[{{Butt} {et~al}\mbox{.}(2006){Butt}, {Drake}, {Benaglia}, {Combi},
  {Dame}, {Miniati}, \& {Romero}}]{bdb+06}
{Butt} Y.~M., {Drake} J., {Benaglia} P., {Combi} J.~A., {Dame} T., {Miniati}
  F., {Romero} G.~E., 2006, ApJ, 643, 238

\bibitem[{{Camilo} {et~al}\mbox{.}(2009){Camilo}, {Ray}, {Ransom}, {Burgay},
  {Johnson}, {Kerr}, {Gotthelf}, {Halpern}, {Reynolds}, {Romani}, {Demorest},
  {Johnston}, {van Straten}, {Saz Parkinson}, {Ziegler}, {Dormody}, {Thompson},
  {Smith}, {Harding}, {Abdo}, {Crawford}, {Freire}, {Keith}, {Kramer},
  {Roberts}, {Weltevrede}, \& {Wood}}]{crr+09}
{Camilo} F. {et~al.}, 2009, ApJ, 705, 1

\bibitem[{{Espinoza} {et~al}\mbox{.}(2011){Espinoza}, {Lyne}, {Stappers}, \&
  {Kramer}}]{elsk11}
{Espinoza} C.~M., {Lyne} A.~G., {Stappers} B.~W., {Kramer} M., 2011, MNRAS,
  414, 1679

\bibitem[{{Hanson}(2003)}]{han03}
{Hanson} M.~M., 2003, ApJ, 597, 957

\bibitem[{{Hobbs} {et~al}\mbox{.}(2010){Hobbs}, {Lyne}, \& {Kramer}}]{hlk10}
{Hobbs} G., {Lyne} A.~G., {Kramer} M., 2010, MNRAS, 402, 1027

\bibitem[{{Hobbs} {et~al}\mbox{.}(2006){Hobbs}, {Edwards}, \&
  {Manchester}}]{hem06}
{Hobbs} G.~B., {Edwards} R.~T., {Manchester} R.~N., 2006, MNRAS, 369, 655

\bibitem[{{Hohle} {et~al}\mbox{.}(2010){Hohle}, {Neuh{\"a}user}, \&
  {Schutz}}]{hns10}
{Hohle} M.~M., {Neuh{\"a}user} R., {Schutz} B.~F., 2010, Astronomische
  Nachrichten, 331, 349

\bibitem[{{Horns} {et~al}\mbox{.}(2007){Horns}, {Hoffmann}, {Santangelo},
  {Aharonian}, \& {Rowell}}]{hhs+07}
{Horns} D., {Hoffmann} A.~I.~D., {Santangelo} A., {Aharonian} F.~A., {Rowell}
  G.~P., 2007, A\&A, 469, L17

\bibitem[{Johnston {et~al}\mbox{.}(1992)Johnston, Manchester, Lyne, Bailes,
  Kaspi, Qiao, \& D'Amico}]{jml+92}
Johnston S., Manchester R.~N., Lyne A.~G., Bailes M., Kaspi V.~M., Qiao G.,
  D'Amico N., 1992, ApJ, 387, L37

\bibitem[{Johnston {et~al}\mbox{.}(1996)Johnston, Manchester, Lyne, D'Amico,
  Bailes, Gaensler, \& Nicastro}]{jml+96}
Johnston S., Manchester R.~N., Lyne A.~G., D'Amico N., Bailes M., Gaensler
  B.~M., Nicastro L., 1996, MNRAS, 279, 1026

\bibitem[{{Kargaltsev} {et~al}\mbox{.}(2014){Kargaltsev}, {Pavlov}, {Durant},
  {Volkov}, \& {Hare}}]{kpd+14}
{Kargaltsev} O., {Pavlov} G.~G., {Durant} M., {Volkov} I., {Hare} J., 2014,
  ApJ, 784, 124

\bibitem[{{Kiminki} {et~al}\mbox{.}(2007){Kiminki}, {Kobulnicky}, {Kinemuchi},
  {Irwin}, {Fryer}, {Berrington}, {Uzpen}, {Monson}, {Pierce}, \&
  {Woosley}}]{kkk+07}
{Kiminki} D.~C. {et~al.}, 2007, ApJ, 664, 1102

\bibitem[{{Lorimer} {et~al}\mbox{.}(2006){Lorimer}, {Faulkner}, {Lyne},
  {Manchester}, {Kramer}, {McLaughlin}, {Hobbs}, {Possenti}, {Stairs},
  {Camilo}, {Burgay}, {D'Amico}, {Corongiu}, \& {Crawford}}]{lfl+06}
{Lorimer} D.~R. {et~al.}, 2006, MNRAS, 372, 777

\bibitem[{{Lyne}(2013)}]{lyn13}
{Lyne} A., 2013, in IAU Symposium, Vol. 291, IAU Symposium, {van Leeuwen} J.,
  ed., pp. 183--188

\bibitem[{{Lyne} {et~al}\mbox{.}(2010){Lyne}, {Hobbs}, {Kramer}, {Stairs}, \&
  {Stappers}}]{lhk+10}
{Lyne} A., {Hobbs} G., {Kramer} M., {Stairs} I., {Stappers} B., 2010, Science,
  329, 408

\bibitem[{{Lyne}(2008)}]{lyn08}
{Lyne} A.~G., 2008, in American Institute of Physics Conference Series, Vol.
  983, 40 Years of Pulsars: Millisecond Pulsars, Magnetars and More, {Bassa}
  C., {Wang} Z., {Cumming} A., {Kaspi} V.~M., eds., pp. 561--566

\bibitem[{{Martins} {et~al}\mbox{.}(2005){Martins}, {Schaerer}, \&
  {Hillier}}]{msh05}
{Martins} F., {Schaerer} D., {Hillier} D.~J., 2005, A\&A, 436, 1049

\bibitem[{{Massey} \& {Thompson}(1991)}]{mt91}
{Massey} P., {Thompson} A.~B., 1991, AJ, 101, 1408

\bibitem[{{Murakami} {et~al}\mbox{.}(2011){Murakami}, {Kitamoto}, {Kawachi}, \&
  {Nakamori}}]{mkk+11}
{Murakami} H., {Kitamoto} S., {Kawachi} A., {Nakamori} T., 2011, PASJ, 63, 873

\bibitem[{{Rauw} {et~al}\mbox{.}(2014){Rauw}, {Naze}, {Wright}, {Drake},
  {Guarcello}, {Prinja}, {Peck}, {Albacete Colombo}, {Herrero}, {Kobulnicky},
  {Sciortino}, \& {Vink}}]{rnw+14}
{Rauw} G. {et~al.}, 2014, ApJS, in Press

\bibitem[{{Ray} {et~al}\mbox{.}(2011){Ray}, {Kerr}, {Parent}, {Abdo},
  {Guillemot}, {Ransom}, {Rea}, {Wolff}, {Makeev}, {Roberts}, {Camilo},
  {Dormody}, {Freire}, {Grove}, {Gwon}, {Harding}, {Johnston}, {Keith},
  {Kramer}, {Michelson}, {Romani}, {Saz Parkinson}, {Thompson}, {Weltevrede},
  {Wood}, \& {Ziegler}}]{rkp+11}
{Ray} P.~S. {et~al.}, 2011, ApJS, 194, 17

\bibitem[{{Rygl} {et~al}\mbox{.}(2012){Rygl}, {Brunthaler}, {Sanna}, {Menten},
  {Reid}, {van Langevelde}, {Honma}, {Torstensson}, \& {Fujisawa}}]{rbs+12}
{Rygl} K.~L.~J. {et~al.}, 2012, A\&A, 539, 79

\bibitem[{{Salas} {et~al}\mbox{.}(2014){Salas}, {Ma{\'{\i}}z Apell{\'a}niz}, \&
  {Barb{\'a}}}]{sab14}
{Salas} J., {Ma{\'{\i}}z Apell{\'a}niz} J., {Barb{\'a}} R.~H., 2014, ArXiv
  e-prints (1410.6767)

\end{thebibliography}

\label{lastpage}
\end{document}